\documentclass[epj,nopacs]{svjour}
\usepackage{latexsym}
\usepackage[square,numbers,sort&compress,nonamebreak]{natbib}
\usepackage{hypernat}
\usepackage{color}
\usepackage{ifpdf}
\usepackage{exscale} 
\usepackage[latin1]{inputenc} 
\usepackage{subfigure}

\newif\ifcomment
\newif\ifprint
\printtrue
\newif\ifbw

\ifpdf
\usepackage[pdftex]{graphicx}
\ifprint
 \usepackage[linkbordercolor={0 0 .8},%
             citebordercolor={.8 0 0},%
             urlbordercolor={.8 0 .8},%
             raiselinks=true,%
             pdfborder={0 0 1 [3]}]{hyperref}
\else
 \usepackage[pdftex,backref,pagebackref,colorlinks,
             linkcolor=blue,citecolor=blue,anchorcolor=blue,
             pagecolor=blue,urlcolor=red,a4paper]{hyperref}
\fi
\pdfoutput=1        
\pdfcatalog{/PageMode(/UseOutlines)} 
\else 
\usepackage[dvips]{graphicx}
\ifprint
 \usepackage[linkbordercolor={0 0 .8},%
             citebordercolor={.8 0 0},%
             urlbordercolor={.8 0 .8},%
             raiselinks=true,%
             pdfborder={0 0 1 [3]}]{hyperref}
\else
 \usepackage[dvips,backref,pagebackref,colorlinks,
             linkcolor=blue,citecolor=blue,anchorcolor=blue,
             pagecolor=blue,urlcolor=red,a4paper]{hyperref}
\fi
\hypersetup{
  pdfauthor   = {Constantin Loizides},
  pdftitle    = {},
  pdfcreator  = {},
  pdfproducer = {},
  pdfsubject  = {},
  pdfkeywords = {}
 }
\fi

\graphicspath{{./pics/}}

\def\noit {}

\newcommand {\snn}       {\sqrt{s_{\scriptscriptstyle{{\rm NN}}}}}

\newcommand {\av}[1]     {\ensuremath{\left< #1 \right>}}

\newcommand {\lsim}      {\,{\buildrel < \over {_\sim}}\,}
\newcommand {\gsim}      {\,{\buildrel > \over {_\sim}}\,}
\newcommand {\hrefurl}[1]{\href{#1}{\url{#1}}}
\newcommand {\arxiv}[1]  {\href{http://www.arxiv.org/#1}{\mbox{#1}}}
\newcommand {\Ref}[1]    {Ref.~\cite{#1}}

\newcommand {\eq}[1]     {eq.~(\ref{#1})}

\newcommand {\Fig}[1]    {Fig.~\ref{#1}}

\newcommand {\mrm}   {\mathrm}
\newcommand {\pt}    {\ensuremath{p_{\mathrm{t}}}}
\newcommand {\pta}   {\ensuremath{p^{\rm assoc}_{\rm t}}}
\newcommand {\ptt}   {\ensuremath{p^{\rm trig}_{\rm t}}}
\newcommand {\dd}    {\mrm{d}}
\newcommand {\RAA}   {R_{\rm AA}}

\newcommand {\IAAa}  {I^{\rm away}_{\rm AA}}
\newcommand {\Ncoll} {\ensuremath{N_{\rm coll}}}

\newcommand {\qhat}  {\ensuremath{\hat{q}}}
\newcommand {\avq}   {\ensuremath{\av{\hat{q}}}}
\newcommand {\gev}   {\mbox{${\rm GeV}$}}
\newcommand {\fm}    {\mbox{${\rm fm}$}}
\newcommand {\qun}   {\gev^{2}/\fm}

\newcommand{\AuAu} {\mbox{Au+Au}}
\newcommand{\CuCu} {\mbox{Cu+Cu}}
\newcommand{\dAu}  {\mbox{d+Au}}

\renewcommand{\AA} {\mbox{nucleus--nucleus}}
\newcommand{\pp}   {\mbox{p+p}}

\begin{document}
\title{High transverse momentum suppression and surface effects\\ in
\CuCu\ and \AuAu\ collisions within the PQM model}
\titlerunning{High transverse momentum suppression and surface effects in \CuCu\ and \AuAu\ collisions}
\author{Constantin Loizides\inst{1,}\inst{2}}
%
%
\institute{Massachusetts Institute of Technology, 77 Massachusetts Avenue, Cambridge, MA 02139, USA \and 
           Email: loizides@mit.edu}
\date{Received: \today\ / Revised version: \today}
%
\abstract{
We study parton suppression effects in heavy-ion collisions within the 
Parton Quenching Model (PQM). After a brief summary of the main features
of the model, we present comparisons of calculations for the nuclear 
modification and the away-side suppression factor to data in \AuAu\ and 
\CuCu\ collisions at $\snn=200~\gev$. We discuss properties of light hadron 
probes and their sensitivity to the medium density within the PQM Monte Carlo 
framework.
\PACS{
      {PACS-key}{discribing text of that key}   \and
      {PACS-key}{discribing text of that key}
     } 
} 
\maketitle

\section{\label{cl:intro}Introduction}
One of the very early and very exciting findings from the experiments running 
at the Relativistic Heavy Ion Collider (RHIC) was the observation of apparent 
jet suppression. At the top RHIC energy, $\snn=200~\gev$,
the mid-rapidity yield of high transverse momentum 
leading particles in \AuAu\ collisions is about a factor of five lower than expected 
from the measurements in \pp\ collisions at the same energy~\cite{cl:phenixRAA,cl:starRAA}. 
Similarly, jet-like correlations on the azimuthally-opposite (`away') side of 
a high-$\pt$ trigger particle are suppressed by a factor of four to five, 
while the near-side correlation strength is almost unchanged~\cite{cl:starIAA}.
The observed suppression persists also at lower center-of-mass energies~\cite{cl:na57RCP}, 
as well as in \CuCu\ collisions at 62.4 and 200 GeV~\cite{cl:phobosRAA}.
The absence of these effects in \dAu\ collisions at $\snn=200~\gev$~\cite{cl:dAu}
strongly supports the picture of partonic energy loss, where energetic partons, 
produced in initial hard scattering processes, lose energy as a consequence of 
the final-state interaction with the dense partonic matter created in \AA\
collisions. The dominant contribution to the energy loss is believed to originate 
from medium-induced gluon radiation (see \Ref{cl:eloss} and references therein). 
Recent calculations point out, however, that the collisional contribution 
to the energy loss might not be negligible~\cite{cl:collloss}. Still, strong interest 
in these probes arises mainly from the fact that modification of their properties 
due to interaction with the medium provides access to fundamental properties of the 
created matter, such as its density and nature~\cite{cl:modelsdata}.

A quite simple model that includes final-state gluon radiation is the Parton 
Quenching Model (PQM)~\cite{cl:PQM}. It combines the pQCD \mbox{BDMPS-Z-SW} framework 
for the probabilistic calculation of parton energy loss in extended partonic matter 
of given size and density~\cite{cl:carlosurs} with a realistic description of the 
collision overlap geometry in a static medium. High-energy partons (and parton pairs) 
are treated on an event-by-event basis using Monte Carlo techniques. 
The model has one single free parameter that sets the scale of the medium transport 
coefficient $\hat{q}$, the average transverse momentum squared transferred to the hard 
parton per unit path length, and, thus, the scale of the energy loss.

In these proceedings, we will, after a short introduction to the PQM model,
compare the suppression phenomena introduced above to the calculations obtained 
with PQM. In particular, we will discuss to what extent light hadronic 
probes are sensitive to the dense matter formed at RHIC.

\section{\label{cl:pqmmodel}The PQM model}
The Monte Carlo calculation of the unquenched and quenched transverse 
momentum spectra in PQM consists of four main steps: 
1) Determination of a parton type and its transverse momentum 
according to PYTHIA (LO) parton distribution functions. 2) Determination 
of its parton-production point in the transverse plane according to the 
nuclear density profile (Glauber) and evaluation of path length and 
transport coefficient seen by the produced parton using two line integrals
weighted with $\rho_{\rm coll}$. 
3) Calculation of the energy loss using constrained quenching weights 
to extrapolate from the eikonal approach used in the BDMPS-SW framework 
to finite parton energies. Two types of constraints are constructed to
estimate the systematic uncertainty of the approach: In the reweighted 
case the energy-loss distribution is simply normalized with-in the 
kinematically allowed regime, whereas in the non-reweighted case 
the fraction of the distribution larger than the energy of the 
parton contributes to the probability of maximum energy loss. 
4) Finally, independent fragmentation is applied to the quenched 
and original parton. Back-to-back parton pairs initially consist of a pair of 
partons with the same $\pt$ at the same production point, but with opposite-side 
emission angle. 

Currently, the model is restricted to mid-rapidity, and, for simplicity, 
we ignore initial-state effects. The single parameter of the model~($k$) 
is fixed to set the scale of the energy loss by fitting to data from  
0--10\% central \AuAu\ collisions at $\snn=200~\gev$. Once the scale 
is fixed we implicitly vary the medium density by its dependence on 
the centrality as given by Glauber. 

Details on the quenching procedure and its application to high-$\pt$ data 
can be found in \Ref{cl:PQM}.

\section{\label{cl:raasuppression}Suppression of leading particles}
Suppression of leading particles is usually quantified via the 
nuclear modification factor,
\begin{equation}      
 \label{cl:eq:raa}
 R_{\rm AA}(\pt,\eta) \equiv 
 \frac{1}{\av{\Ncoll}} \times
 \frac{{\rm d}^2 N_{\rm AA}/{\rm d}\pt{\rm d}\eta}
 {{\rm d}^2 N_{\rm pp}/{\rm d}\pt{\rm d}\eta} \,,  
\end{equation}
the ratio of the yield of light hadrons in \AA\ over proton--proton 
yield scaled by the number of binary collisions in a given centrality class. 
The ratio is normalized so that, if no final-state effects were 
present, it would be close to one.  
Indeed, this scaling is observed in the measurement of direct photon 
yield in \AuAu\ collisions at $\snn=200~\gev$~\cite{cl:photons}.

\begin{figure}[b]
\includegraphics[width=0.49\textwidth]{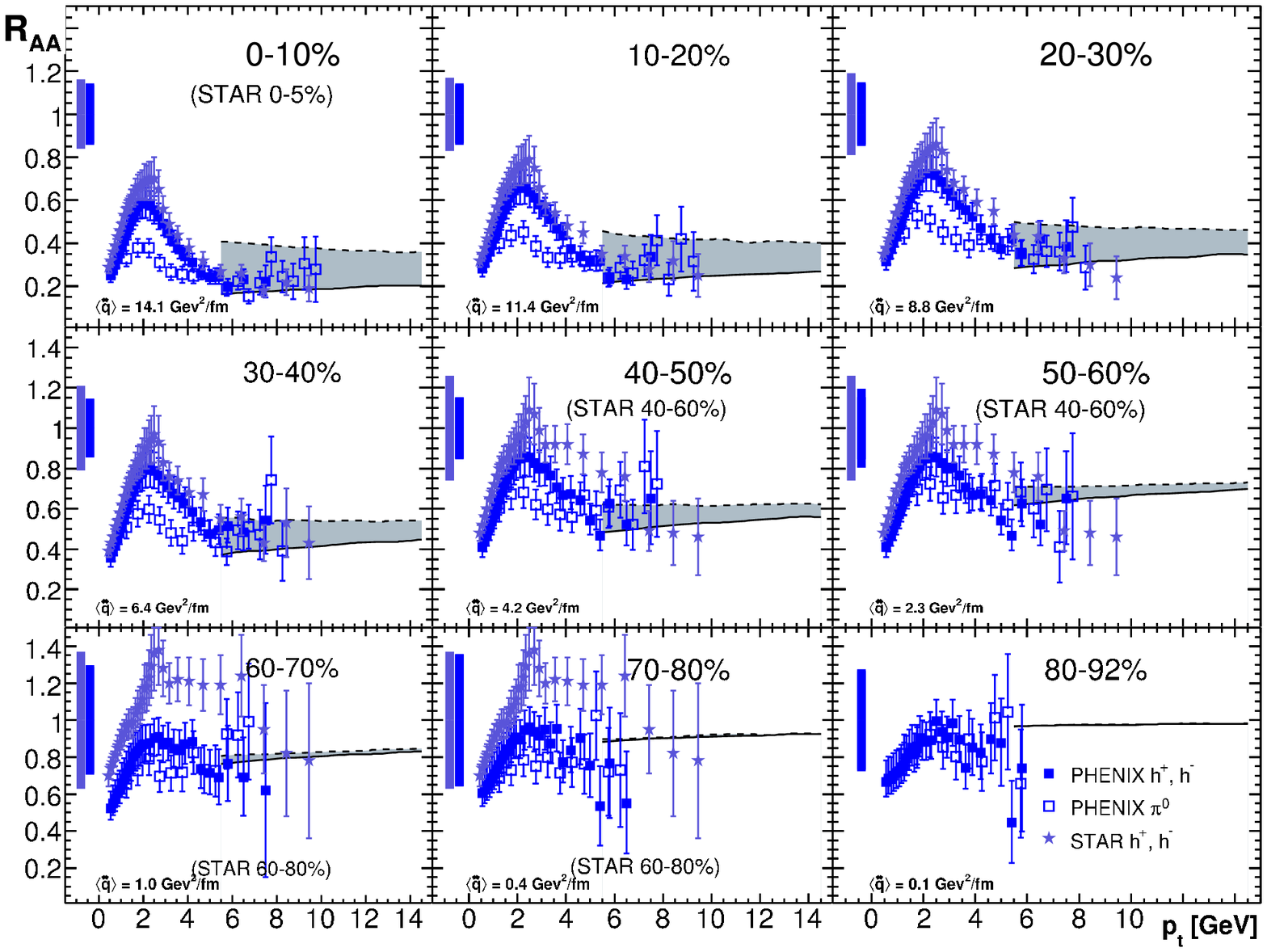}
\caption[]{\label{cl:fig1}
 $\RAA(\pt)$ in \AuAu\ at $\snn=200~\gev$ for different centralities.
 Data are PHENIX charged hadrons and $\pi^0$~\cite{cl:phenixRAA}
 and STAR charged hadrons~\cite{cl:starRAA} with combined
 statistical and $\pt$-dependent systematic errors (bars on the 
 data points) and $\pt$-independent systematic errors (bars at 
 \mbox{$\RAA=1$}). The gray band is the original PQM calculation 
 from \Ref{cl:PQM} using reweighted and non-reweighted quenching
 weights.
}
\end{figure}

However, in \AuAu\ collisions at the same energy, $\RAA$ for light hadrons 
at mid-rapidity is found to decrease from peripheral ($\RAA\simeq 1$) 
to central events ($\RAA\simeq 0.2$), for $\pt\gsim 5~\gev$~(see \Fig{cl:fig1}). 
In this high-$\pt$ region, $\RAA$ is independent of the particle type and rather 
independent of $\pt$. In \Fig{cl:fig1}, the data is compared to the original 
calculation of PQM, where the single parameter of the model was adjusted to match 
the suppression measured in 0--10\% central \AuAu\ collisions at $\snn=200~\gev$. 
For central collisions, this corresponds to an average transport coefficient 
of about $\av{\hat{q}}=14~\gev^2/\fm$, which decreases with decreasing centrality
to essentially zero for the most peripheral collisions. The average is taken over all 
produced hard partons and given in the equivalent static scenario~\cite{cl:carlosurs}.

\begin{figure*}
\begin{minipage}[t]{0.49\textwidth} 
\includegraphics[width=0.99\textwidth]{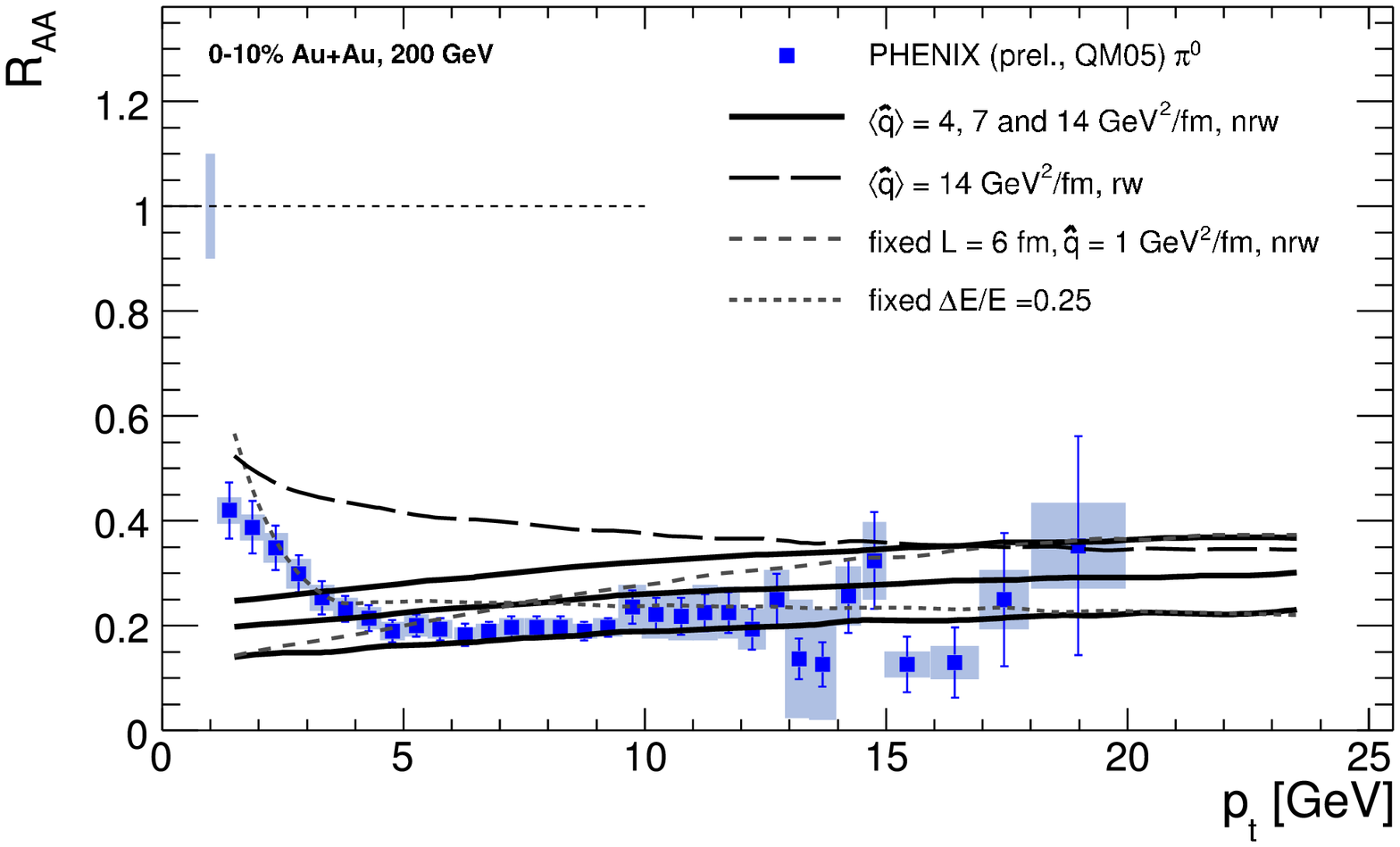}
\caption[]{\label{cl:fig2}
 $\RAA(\pt)$ for neutral pions in 0--10\% central \AuAu\ collisions at 
 \mbox{$\snn=200~\gev$} (preliminary PHENIX data) with combined statistical 
 and $\pt$-dependent systematic errors, as well as $\pt$-independent systematic 
 errors~(bar at \mbox{$\RAA=1$})~\cite{cl:phenixPrelimRAA}. 
 The PQM curves for $\avq=14~\qun$ are the original PQM results (extended to 
 larger $\pt$) for the reweighted and non-reweighted case from \Ref{cl:PQM}. 
 In addition, results for $\avq=4$ and 
 $7~\qun$ for the non-reweighted case are shown, as well as calculations 
 for fixed $L=6~\fm$ and $\qhat=1~\qun$, or for fixed relative energy 
 loss of $\Delta E/E=0.25$.
}
\end{minipage}
\hspace{0.2cm} 
\begin{minipage}[t]{0.49\textwidth}
\includegraphics[width=0.99\textwidth]{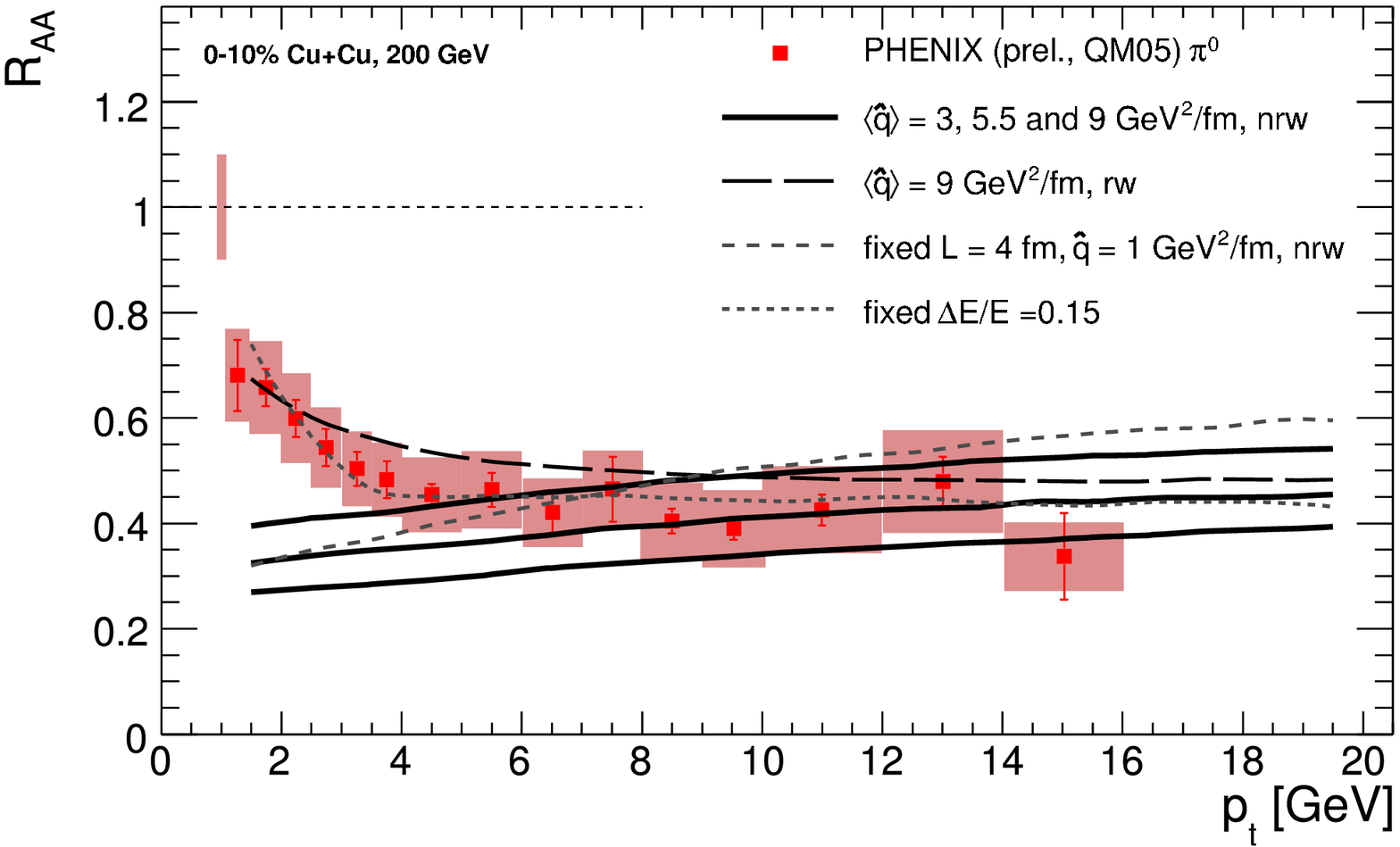}
\caption[]{\label{cl:fig3}
 $\RAA(\pt)$ for neutral pions in  0--10\% central \CuCu\ collisions at 
 \mbox{$\snn=200~\gev$} (preliminary PHENIX data) with combined statistical and 
 $\pt$-dependent systematic errors, as well as $\pt$-independent systematic 
 errors~(bar at \mbox{$\RAA=1$})~\cite{cl:phenixPrelimRAA}. 
 The PQM curves for $\avq=9~\qun$ are the original PQM predictions for the reweighted 
 and non-reweighted case from \Ref{cl:PQM}. In addition, results for $\avq=3$ and 
 $5.5~\qun$ for the non-reweighted case are shown, as well as calculations 
 for fixed $L=4~\fm$ and $\qhat=1~\qun$, or for fixed relative energy loss of 
 $\Delta E/E=0.15$.
}
\end{minipage}
\vfill
\vspace{0.5cm}
\begin{minipage}[t]{0.49\textwidth} 
\includegraphics[width=0.99\textwidth]{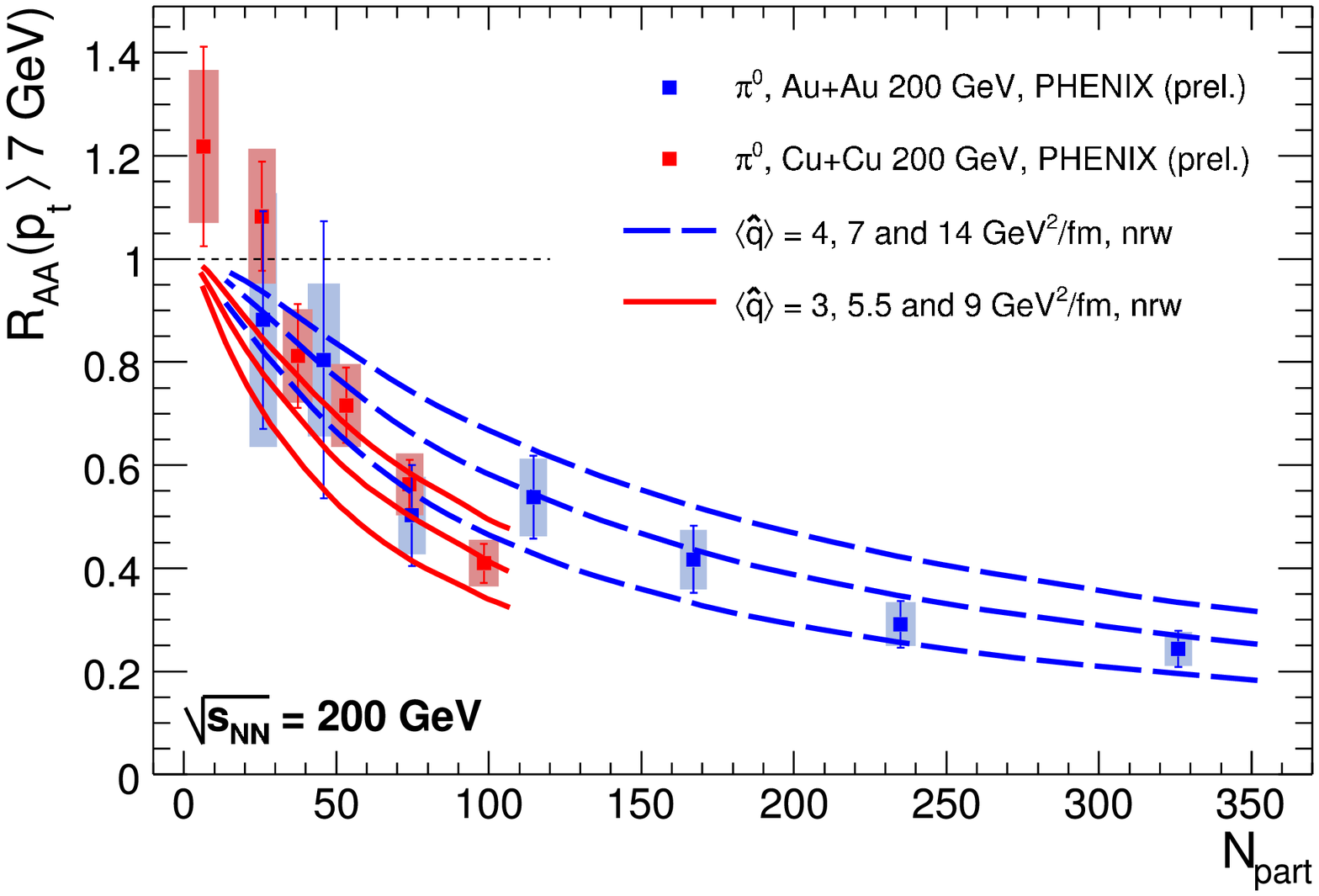}
\caption[]{\label{cl:fig4}
 Centrality dependence of $\RAA(\pt>7~\gev)$ for neutral pions in \CuCu\ and 
 \AuAu\ collisions at \mbox{$\snn=200~\gev$} (preliminary PHENIX data) with 
 combined statistical and systematic errors~\cite{cl:phenixPrelimRAA}.
 The PQM calculation is shown for $\avq=4$, $7$ and $14~\qun$ (\AuAu), as well
 as for $\avq=3$, $5.5$ and $9~\qun$ (\CuCu) in the non-reweighted case.
}
\end{minipage}
\hspace{0.2cm} 
\begin{minipage}[t]{0.49\textwidth}
\includegraphics[width=0.99\textwidth]{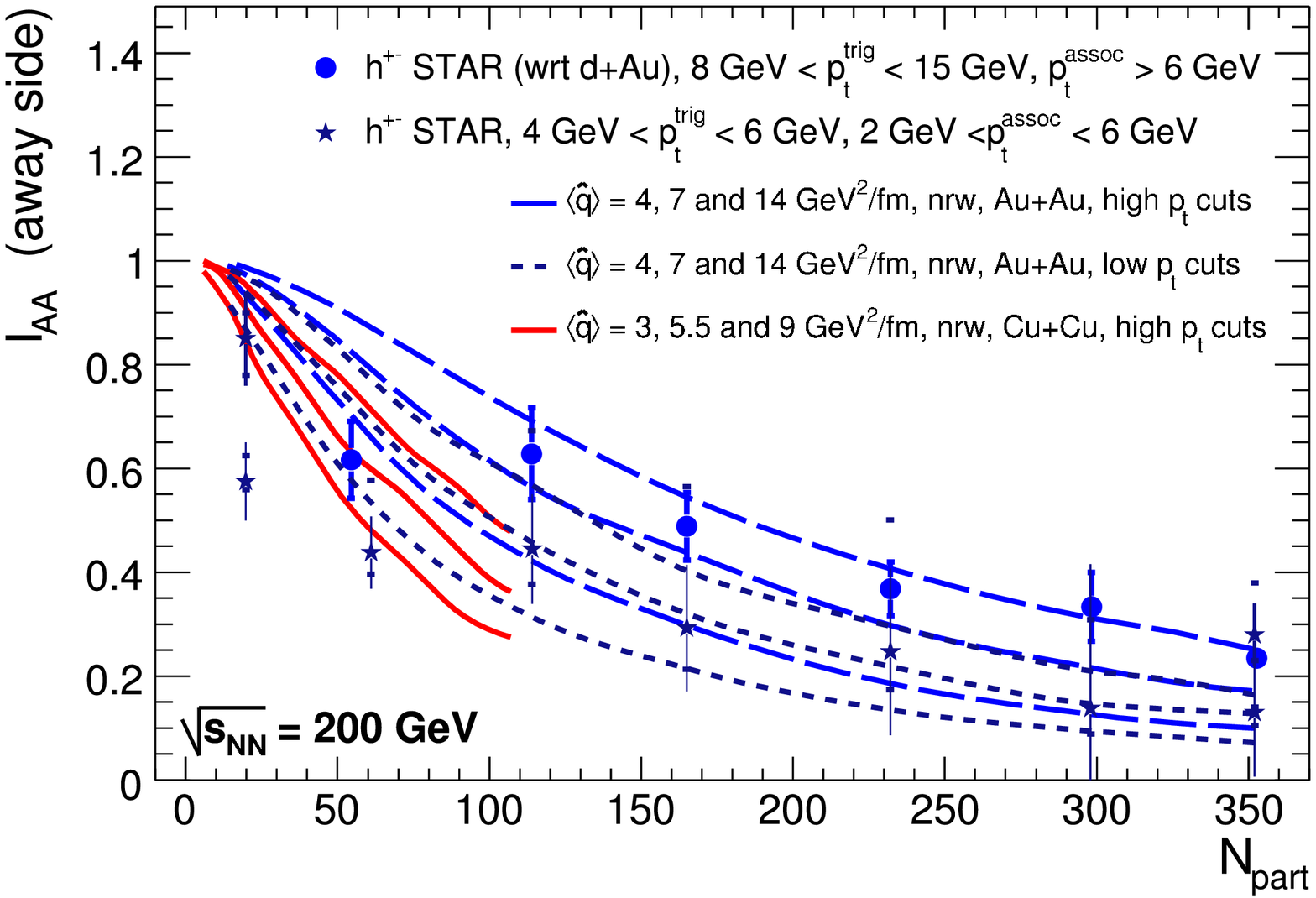}
\caption[]{\label{cl:fig5}
 Centrality dependence of $\IAAa$ in \AuAu\ at $\snn=200~\gev$ for two data sets.
 The data for the low $\pt$ cuts of $4<\ptt\le6~\gev$ and $2~\gev\le\pta\le\ptt$ 
 with statistical (bars) and systematic (ticks) errors are from \Ref{cl:starIAA}. 
 The data for the high $\pt$ cuts of $8<\ptt\le15~\gev$ and $6~\gev\le\pta\le\ptt$ 
 with statistical (bars) errors are from \Ref{cl:starDirectJets}~(wrt 
 \dAu\ instead  of pp).
 PQM results in the non-reweighted case for \AuAu\ are for $\avq=4$, 
 $7$ and $14~\qun$, and for \CuCu\ for $\avq=3$, $5.5$ and $9~\qun$ (only high 
 trigger cuts).
}
\end{minipage}
\end{figure*} 

In \Fig{cl:fig2}, we show $\RAA(\pt)$ for neutral pions in 0--10\% central
\AuAu\ collisions at \mbox{$\snn=200~\gev$} for preliminary PHENIX 
data~\cite{cl:phenixPrelimRAA}, which extends beyond the previous data by
almost a factor of two in $\pt$. The new data is compared to the 
original PQM calculation for $\avq=14~\qun$ (simply extended to larger $\pt$) 
in the reweighted and non-reweighted approximation of the quenching
weights for finite (small) parton energies. 
For a given medium density, the reweighted values generally produce 
a weaker energy loss than the non-reweighted ones. However,
in the reweighted case, only a rather small fraction of the full 
quenching weights is used. 
This results in the unphysical behaviour that, beyond a certain medium 
density, the probability of `zero energy loss' increases for increasing 
densities.
Therefore, we show in addition, results for lower densities, for $\avq=4$ 
and $7~\qun$, in the non-reweighted case. We furthermore compare the
realistic calculations with calculations where we either fix the
geometry ($L=6~\fm$ and $\qhat=1~\qun$) or the relative energy loss
($\Delta E/E=0.25$). While the fixed geometry case leads to a more 
strongly rising $\RAA$ with increasing $\pt$, the fixed relative energy 
loss approximation over a very wide range of $\pt$ describes the data.

Similar conclusions can be made for the comparison with data from \CuCu\ 
interactions at RHIC. In \Fig{cl:fig3}, we show $\RAA(\pt)$ for neutral pions
in 0--10\% central \CuCu\ collisions at \mbox{$\snn=200~\gev$} for preliminary 
PHENIX data~\cite{cl:phenixPrelimRAA}. The data are compared to the PQM 
prediction (made before QM'05~\cite{cl:preQMprediction}) for 
$\avq=9~\qun$ for the reweighted and non-reweighted case. The value of
$\avq=9~\qun$ relies on the proportionality of the transport coefficient 
to the initial volume-density of gluons~\cite{cl:baier} and on the 
predictions of the saturation model~\cite{cl:ekrt}, as outlined in 
\Ref{cl:PQM}.
The prediction seems to slightly overestimate the suppression in the smaller
\CuCu\ system. We therefore also present results for lower densities, 
for $\avq=3$ and $5.5~\qun$, in the non-reweighted case. As before, the 
calculations including the nuclear geometry are compared to calculations 
with either fixed geometry ($L=4~\fm$ and $\qhat=1~\qun$) or relative 
energy loss ($\Delta E/E=0.15$). 

In \Fig{cl:fig4}, we show calculations of the centrality dependence of 
the nuclear modification factor for $\pt>7~\gev$ in \CuCu\ and \AuAu\ 
collisions at \mbox{$\snn=200~\gev$} compared to the preliminary PHENIX 
data~\cite{cl:phenixPrelimRAA}. The data seem to favour values for the 
transport coefficient of $\avq=7-14$ for \AuAu\ and $3-5.5~\qun$ for 
\CuCu\ collisions.

\section{\label{cl:iaasuppression}Suppression of jet-like correlations}
Within the PQM framework, we consider the suppression of back-to-back 
jet-like correlations by simulation of back-to-back pairs of partons 
in a simple LO parton model (no intrinsic $k_{\rm t}$).
The magnitude of the suppression is usually quantified 
by the factor $\IAAa = D^{\rm away}_{\rm AA}/D^{\rm away}_{\rm pp}$,
where the di-hadron correlation strength, $D^{\rm away}_{\rm pp(AA)}$,
for an associated hadron, $h_2$, with $p_{\rm t,2}$ in the opposite 
azimuthal direction from a trigger hadron, $h_1$, with $p_{\rm t,1}$,
is integrated over the considered trigger- and associated- 
$\pt$ intervals~\cite{cl:wang},
\begin{eqnarray}
 \label{cl:eq:daapp}
 D^{\rm away}_{\rm pp(AA)} &=&
 \int_{p^{\rm trig}_{\rm t,min}}^{p^{\rm trig}_{\rm t,max}}\dd p_{\rm t,1}
 \int_{p^{\rm assoc}_{\rm t,min}}^{p^{\rm assoc}_{\rm t,max}}\dd p_{\rm t,2}
 \\\nonumber
 & &\int_{\rm{away\, side}}\dd\Delta\phi\,
 \frac{\dd^3\sigma_{\rm pp(AA)}^{h_1h_2}/\dd p_{\rm t,1}\dd p_{\rm t,2}
 \dd\Delta\phi}
 {\dd\sigma_{\rm pp(AA)}^{h_1}/\dd p_{\rm t,1}}\,,
\end{eqnarray}

As was found for the nuclear modification factor in \AuAu\ collisions at 
$\snn=200~\gev$, $\IAAa$ for $4<\ptt\le6~\gev$ and $2~\gev\le\pta\le\ptt$ 
is found to decrease with increasing centrality, down to about $0.2$--$0.3$ 
for the most central events (see \Fig{cl:fig5}). 
In the figure, we also show $\IAAa$ data taken from \Ref{cl:starDirectJets} for 
higher $\pt$ cuts of $8<\ptt\le15~\gev$, where we normalize the away-side yields 
measured in \AuAu\ to the yields measured in \dAu\ collisions. Both data sets 
are compared to PQM calculations (using \pp\ as the reference in both cases) for 
the non-reweighted case with $\avq=4$, $7$ and $14~\qun$ for \AuAu\ collisions. 
The data for the higher trigger cuts favour smaller medium densities, while the 
uncertainties on the data for the lower cuts make it difficult to draw strong 
conclusions. For completeness, we also present predictions for the \CuCu\ system 
with $\avq=3$, $5.5$ and $9~\qun$ (only high trigger cuts).

\section{\label{cl:sensitivity}Sensitivity of light hadronic probes}
\begin{figure}[t]
\includegraphics[width=0.49\textwidth]{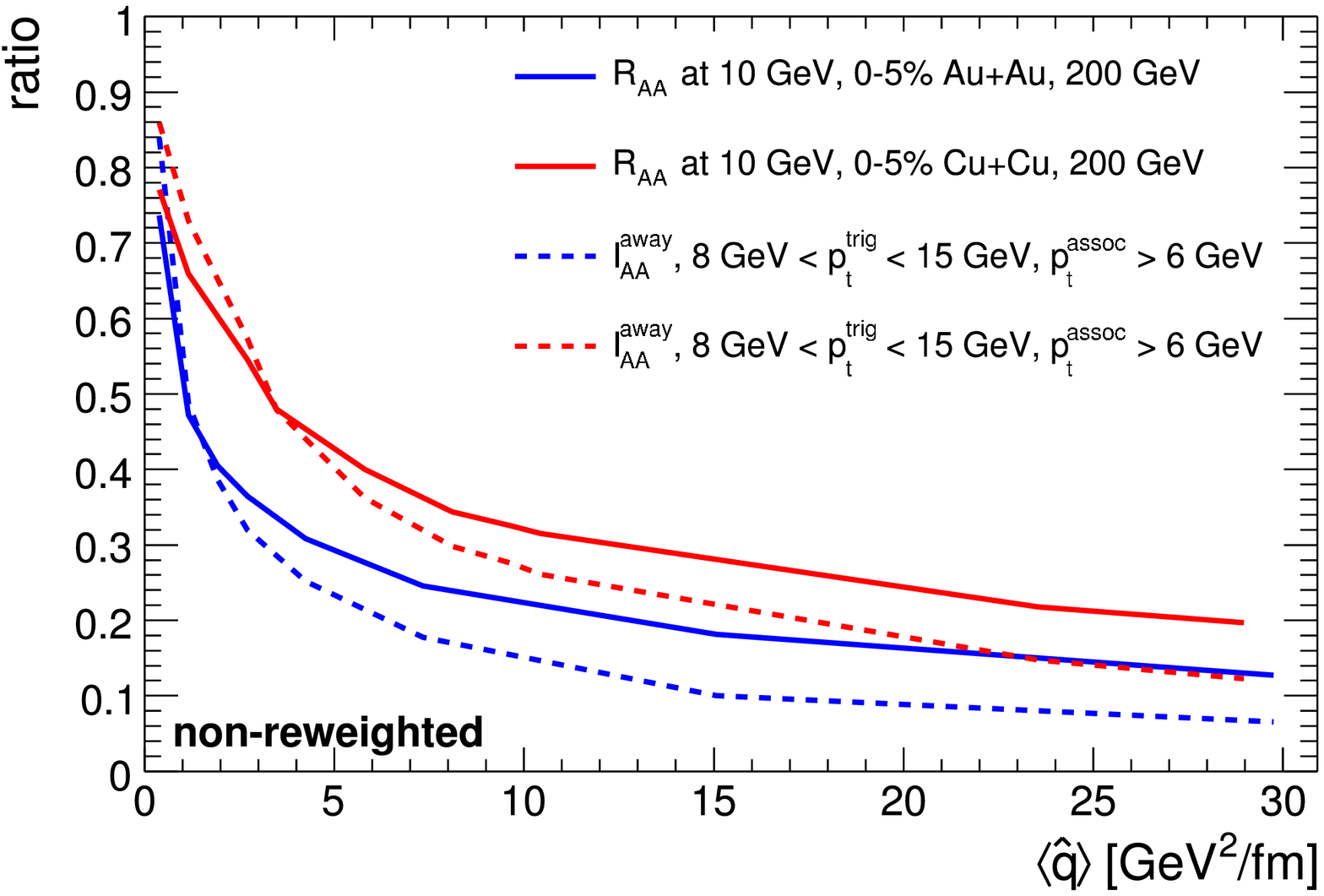}
\caption[]{\label{cl:fig6}
 $\RAA$ at $10~\gev$ and $\IAAa$ for trigger cuts of 
 $8<\ptt\le15~\gev$ and $6~\gev\le\pta\le\ptt$ as a function
 of $\avq$ in 0--5\% central \AuAu\ and \CuCu\ collisions at 
 $\snn=200~\gev$. The calculations are in the non-reweighted 
 approximation.
}
\end{figure}

\enlargethispage{0.5cm}
Another complication arises from the fact that probes based on 
leading-particle analyzes are affected by several biases. The biases
result primarily from the steeply-falling underlying production cross 
sections and the emission from regions close to the surface. The latter
effect dominates for large medium densities. This is illustrated in 
\Fig{cl:fig6}, which shows the behaviour of $\RAA$ (at $10~\gev$) and $\IAAa$ 
(for trigger cuts of $8<\ptt\le15~\gev$ and $6~\gev\le\pta\le\ptt$) as a 
function of the medium density, expressed as $\avq$, in \AuAu\ and \CuCu\ 
collisions at $\snn=200~\gev$. Clearly, beyond a certain medium density,
the numerical values for the ratios saturate at a non-zero value.
The fundamental reason is that the probability of no medium-induced gluon 
radiation, $P(\Delta E=0)$, for a medium of finite size and finite density 
is not zero. The `no radiation' contribution to the spectra is dramatically 
enhanced when realistic nuclear path-length distributions and density profiles 
are taken into account, since $P(\Delta E=0)$ is decreasing as a function 
of $\hat{q}\,L^3$, giving significant weight to partons that `feel' lower 
values of $\hat{q}\,L^3$. We have reported in~\Ref{cl:PQM} that the $\RAA$ 
and $\IAAa$ data (for low trigger cuts) can be described by taking into account 
only $P(\Delta E=0)$ and $1-P(\Delta E=0)$~(see also~\cite{cl:drees}).

Recently, direct measurements of dijets in heavy-ion collisions have been 
performed by the STAR collaboration~\cite{cl:starDirectJets}. It is found that,
while the relative yields of the hadron-triggered fragmentation function 
relative to \dAu, the integrand in \eq{cl:eq:daapp}, are suppressed, the shape 
is not modified, even in the most central \AuAu\ collisions at $200~\gev$. 
Such a scenario naturally follows from the trigger bias since, due to the cuts 
on near-side and away-side particle $\pt$, those dijet pairs that escaped the 
collision region without losing a significant fraction of their 
initial energy are preferentially selected.

\begin{figure*}[p]
\begin{center}

\ifbw
 \includegraphics[width=0.425\textwidth]{cQGEmission-AuAuCent0-10QhatAway14-bw}
 \hspace{0.8cm}
 \includegraphics[width=0.425\textwidth]{cQGEmission-Eps-AuAuCent0-10Qhat14-bw}
 \vfill
 \vspace{0.2cm}
 \includegraphics[width=0.425\textwidth]{cQGEmission-AuAuCent0-10QhatAway7-bw}
 \hspace{0.8cm}
 \includegraphics[width=0.425\textwidth]{cQGEmission-Eps-AuAuCent0-10Qhat7-bw}
 \vfill
 \vspace{0.2cm}
 \includegraphics[width=0.425\textwidth]{cQGEmission-AuAuCent0-10QhatAway4-bw}
 \hspace{0.8cm}
 \includegraphics[width=0.425\textwidth]{cQGEmission-Eps-AuAuCent0-10Qhat4-bw}
\else
 \includegraphics[width=0.425\textwidth]{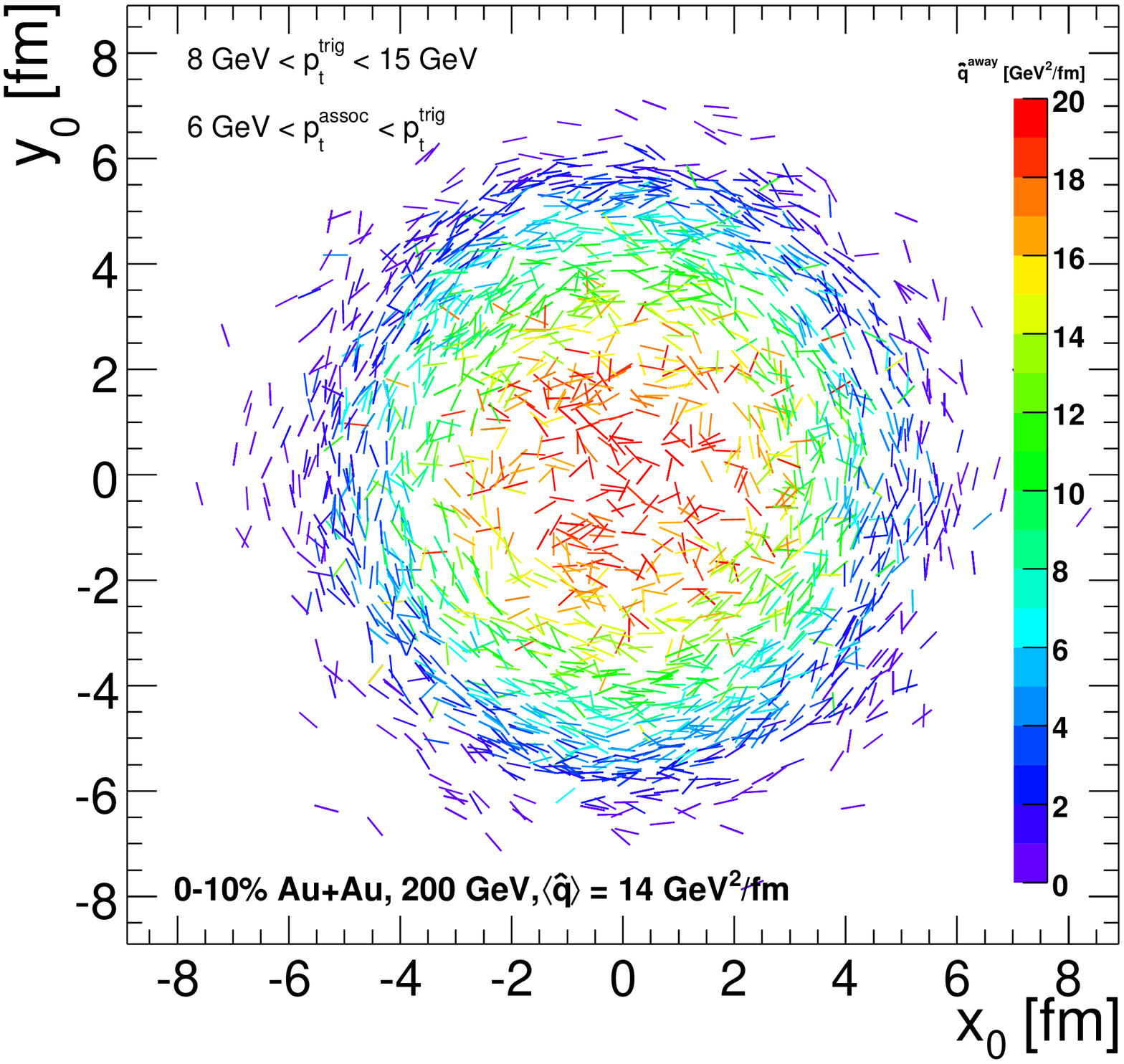}
 \hspace{0.8cm}
 \includegraphics[width=0.425\textwidth]{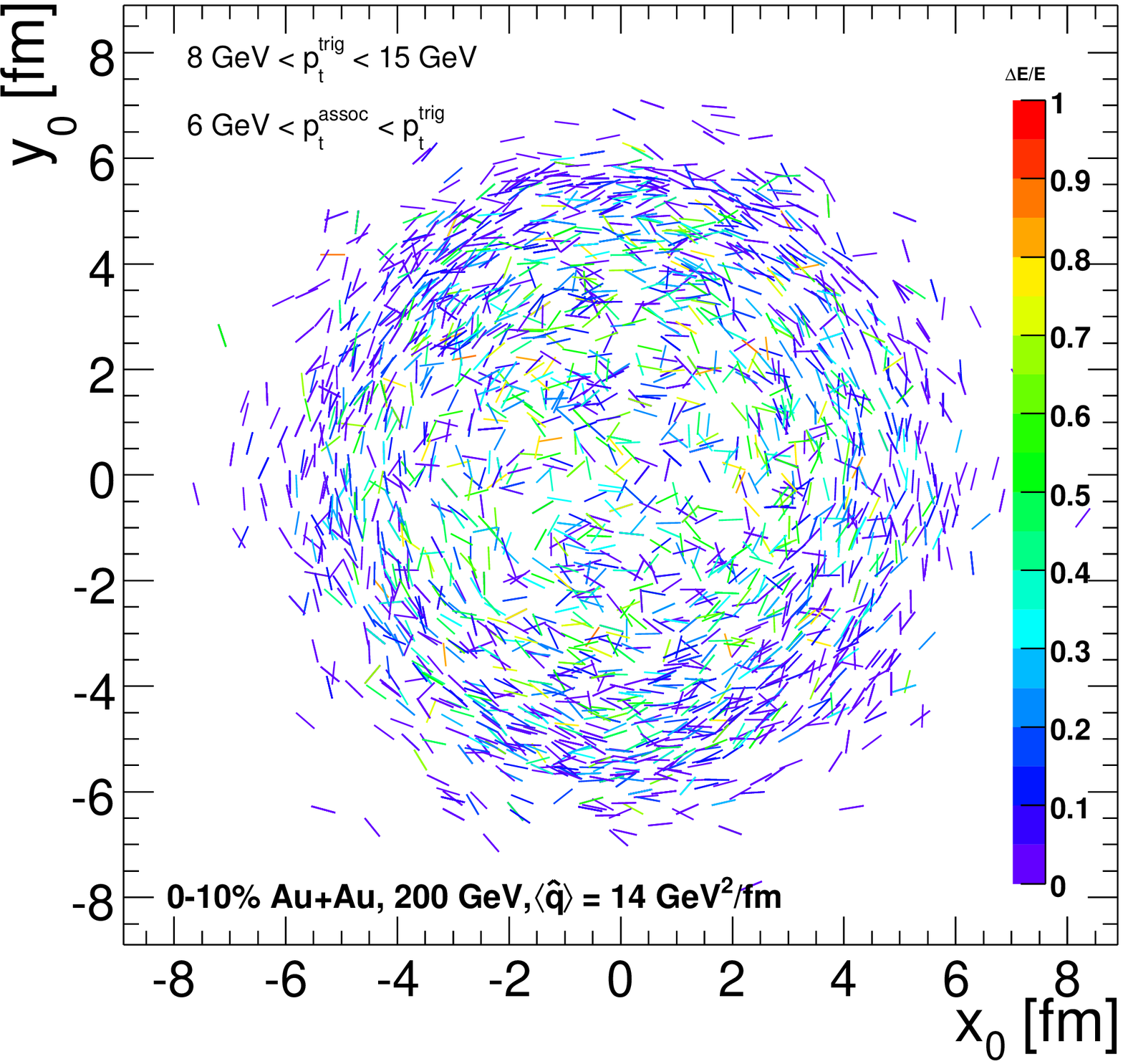}
 \vfill
 \vspace{0.2cm}
 \includegraphics[width=0.425\textwidth]{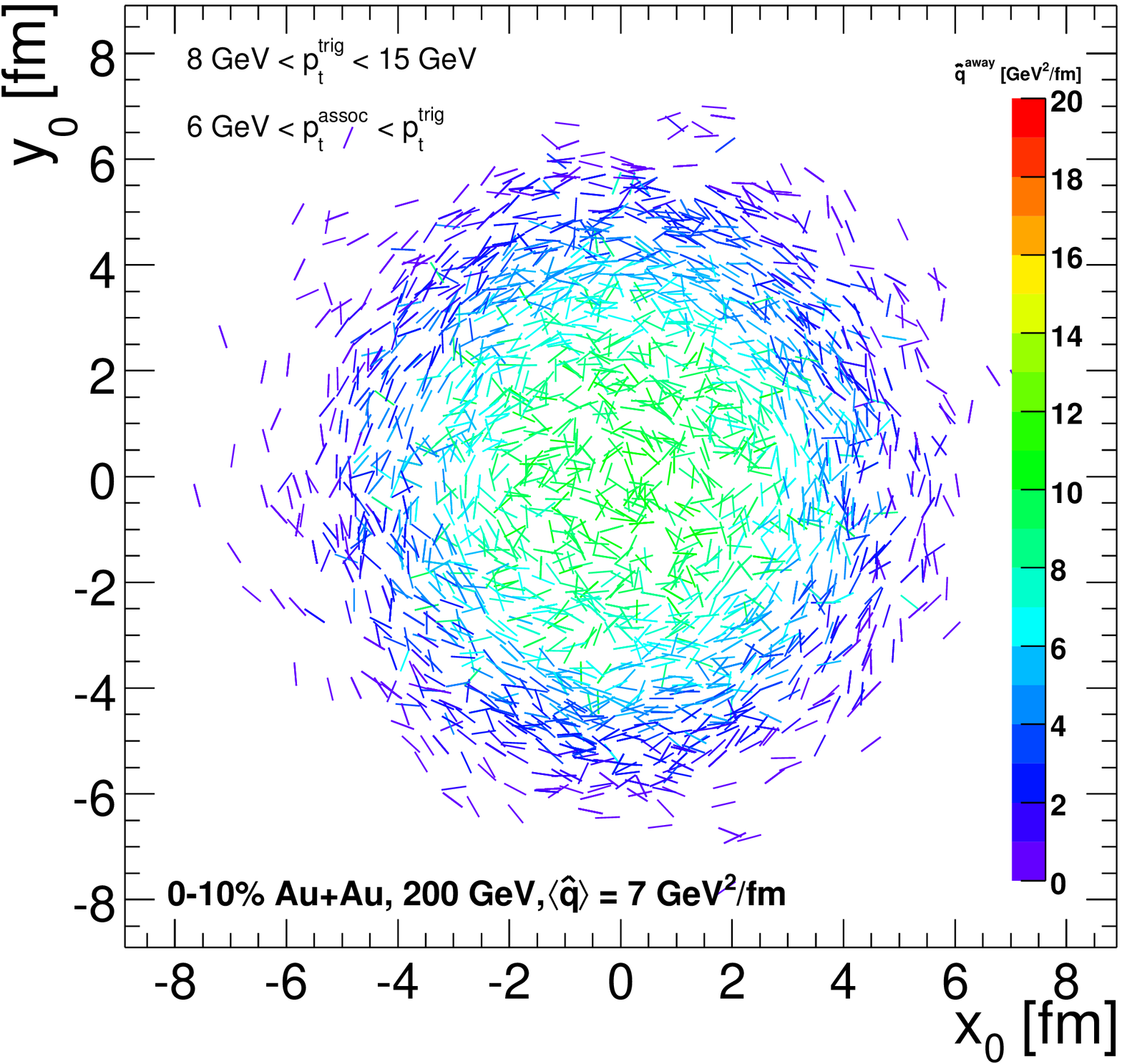}
 \hspace{0.8cm}
 \includegraphics[width=0.425\textwidth]{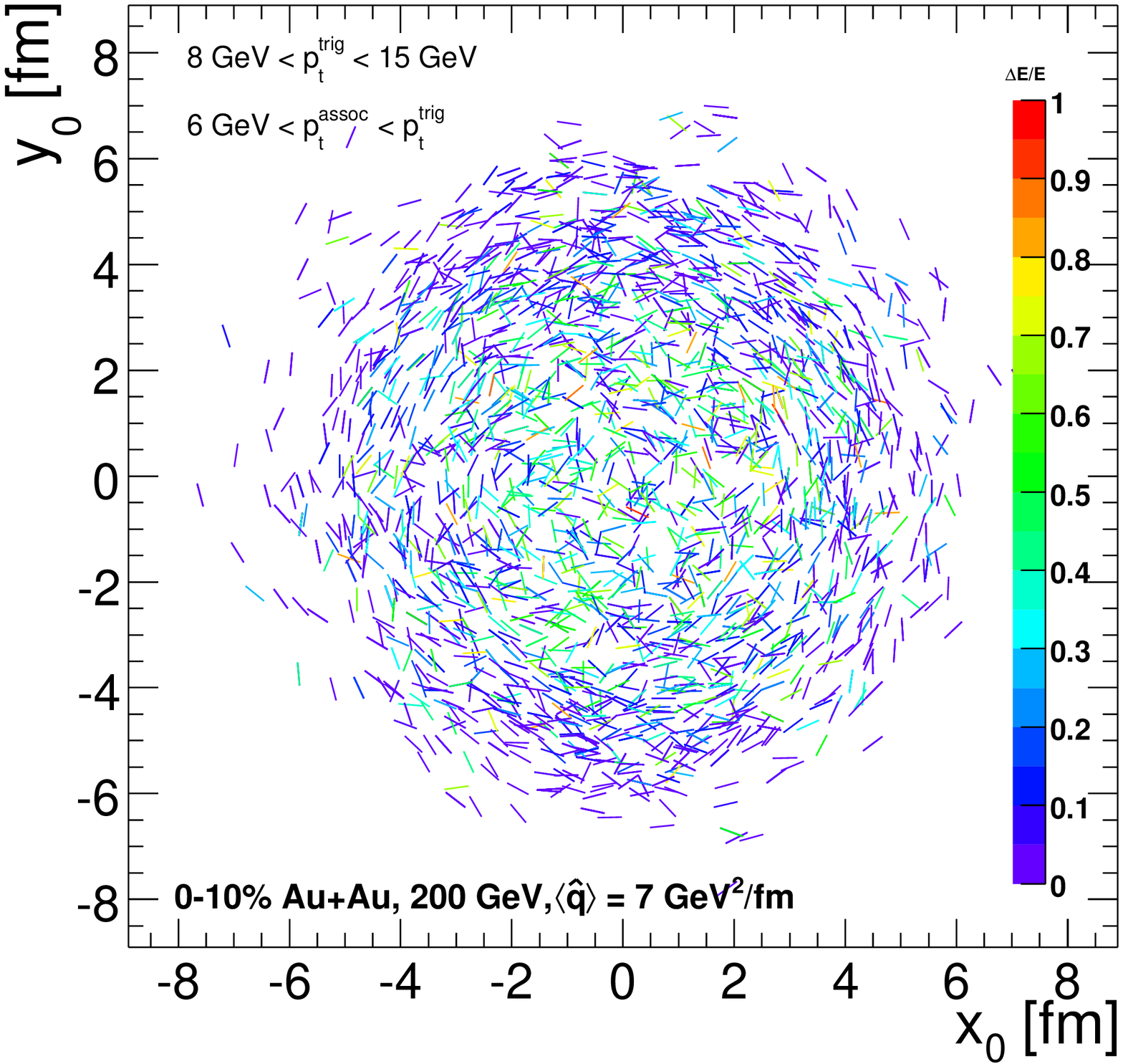}
 \vfill
 \vspace{0.2cm}
 \includegraphics[width=0.425\textwidth]{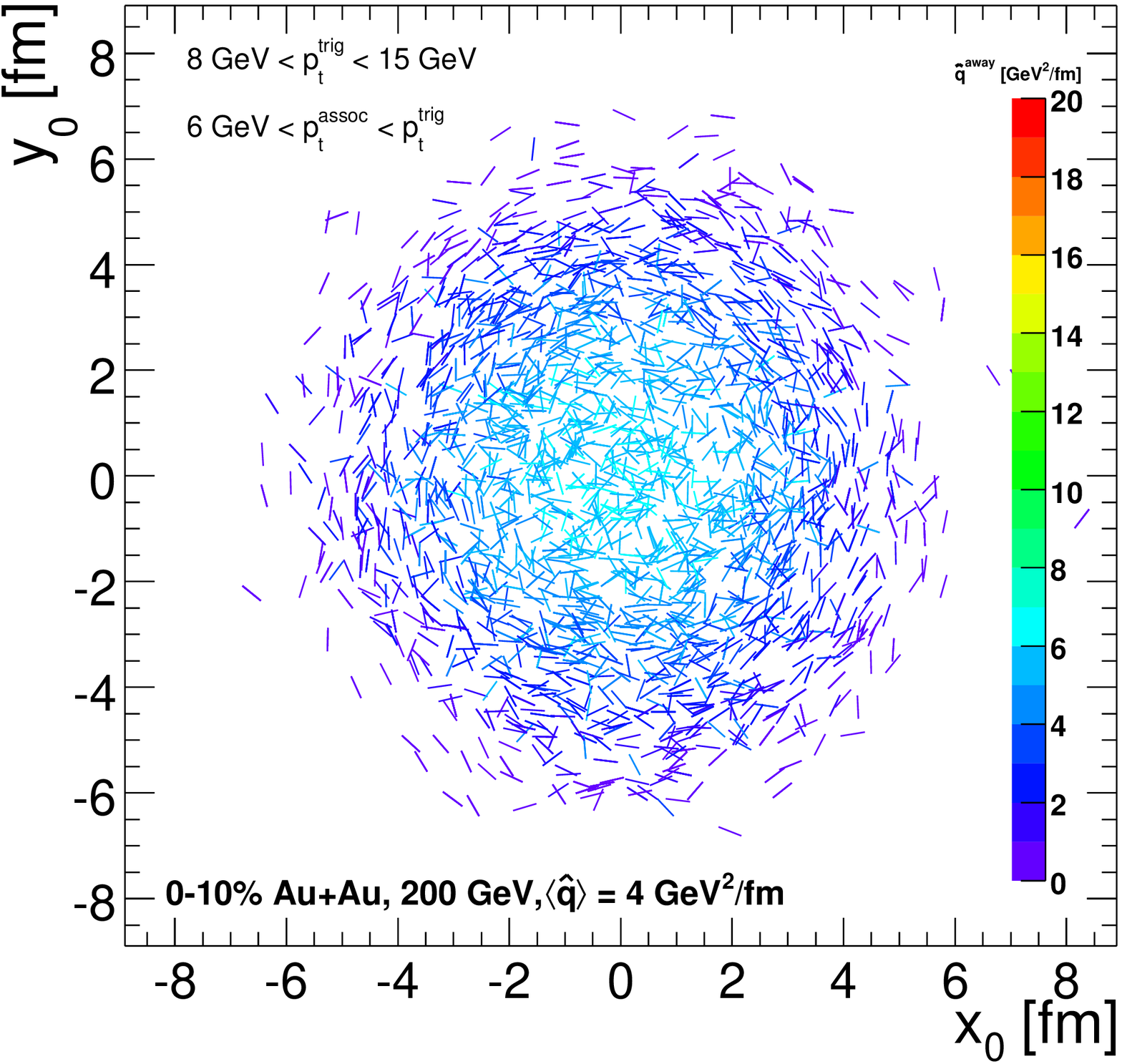}
 \hspace{0.8cm}
 \includegraphics[width=0.425\textwidth]{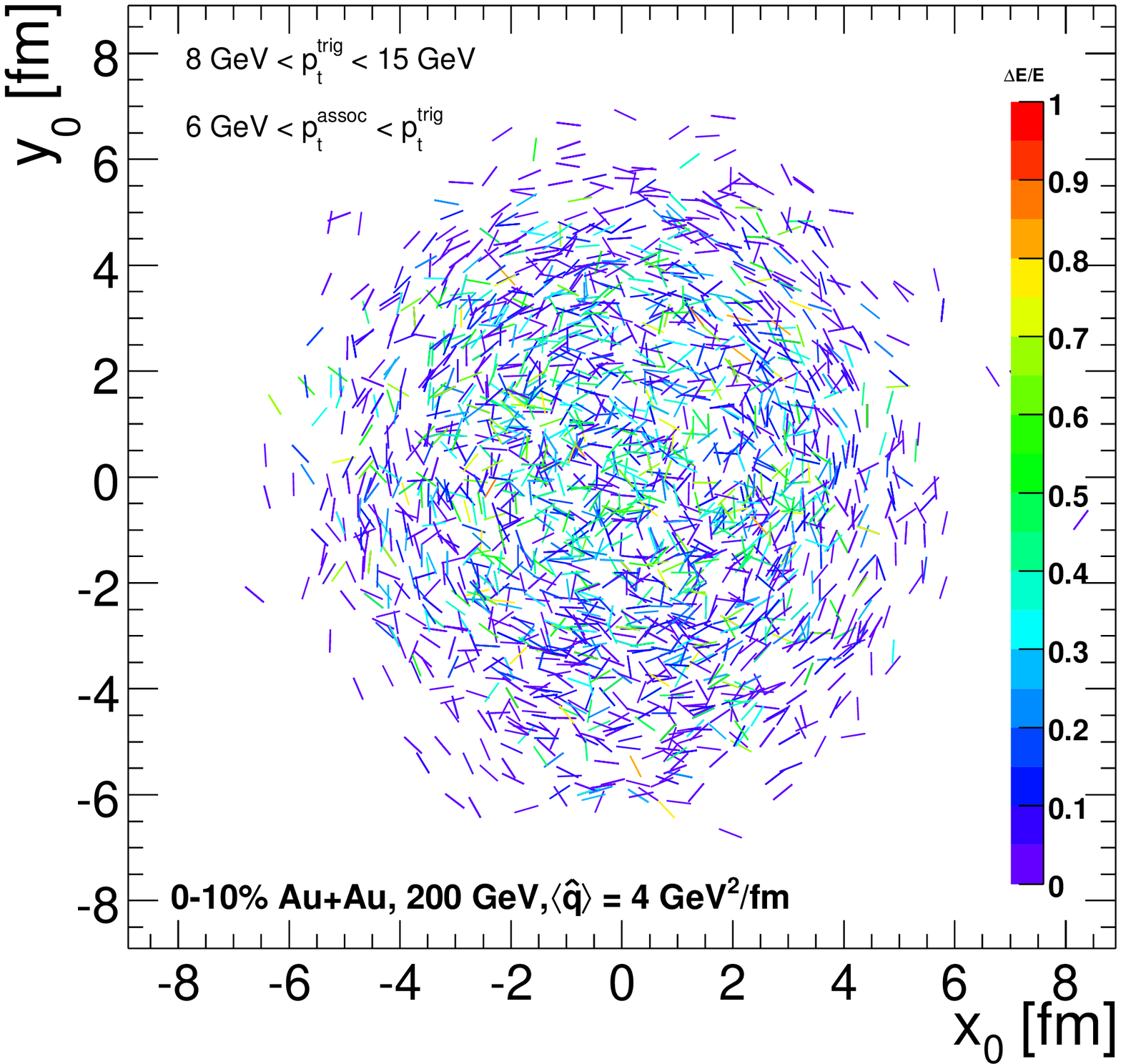}
\fi

\end{center}
\caption[]{\label{cl:fig7} (Color online)
 Production points and emission direction for surviving back-to-back parton 
 pairs (yielding hadron pairs within $8<\ptt\le15~\gev$ and $6~\gev\le\pta\le\ptt$)
 in the transverse plane for $\avq=4$~(bottom), $7$~(middle) and $14~\qun$~(top
 panels) for 0--10\% central \AuAu\ collisions. 
 The line color represents the medium density~(relative energy loss) of the 
 away-side parton in the left~(right) panel.
} 
\end{figure*}

\begin{figure}[t]
\begin{center}
\subfigure[]{\label{cl:fig8a}
\includegraphics[width=0.49\textwidth]{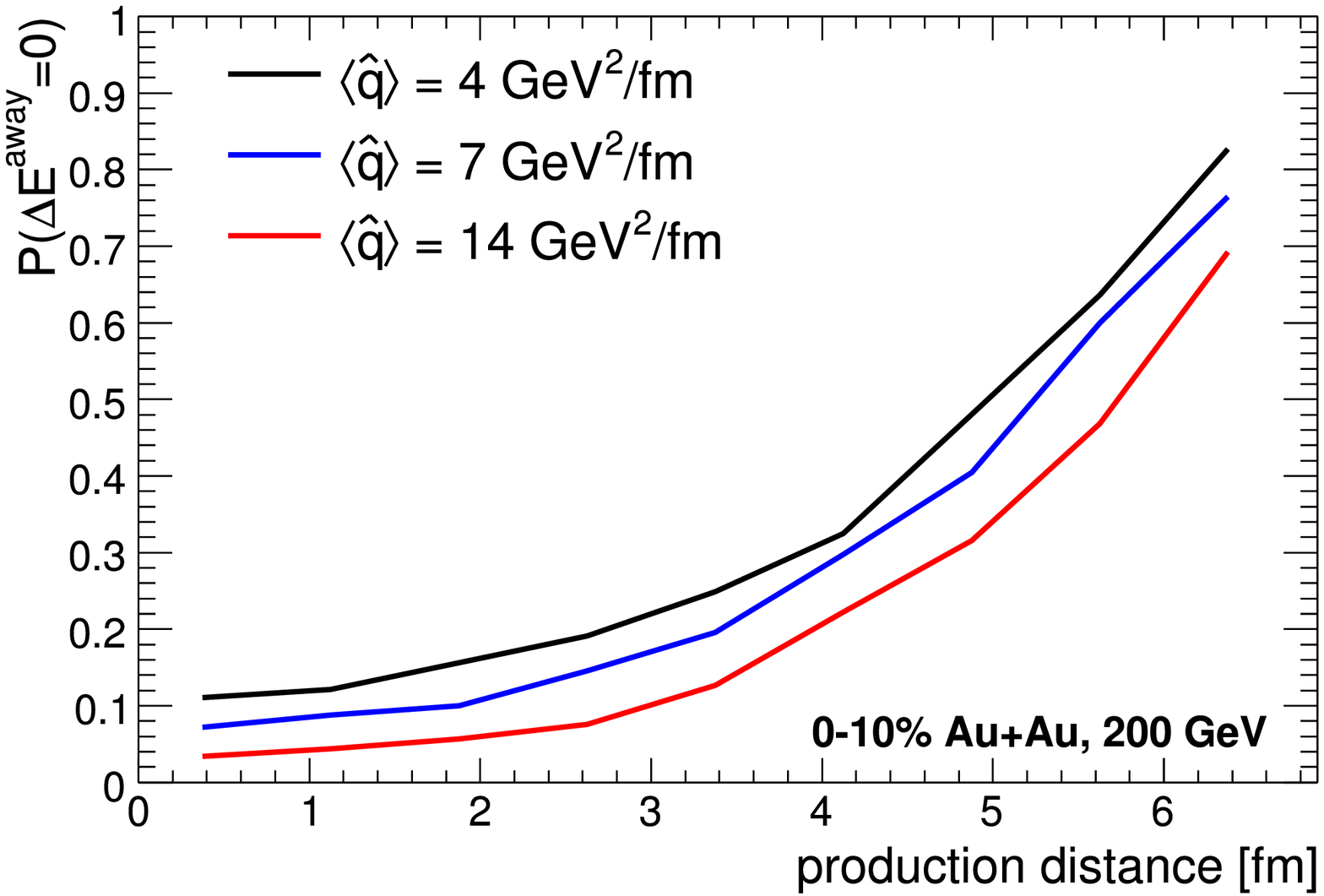}}
\subfigure[]{\label{cl:fig8b}
\includegraphics[width=0.49\textwidth]{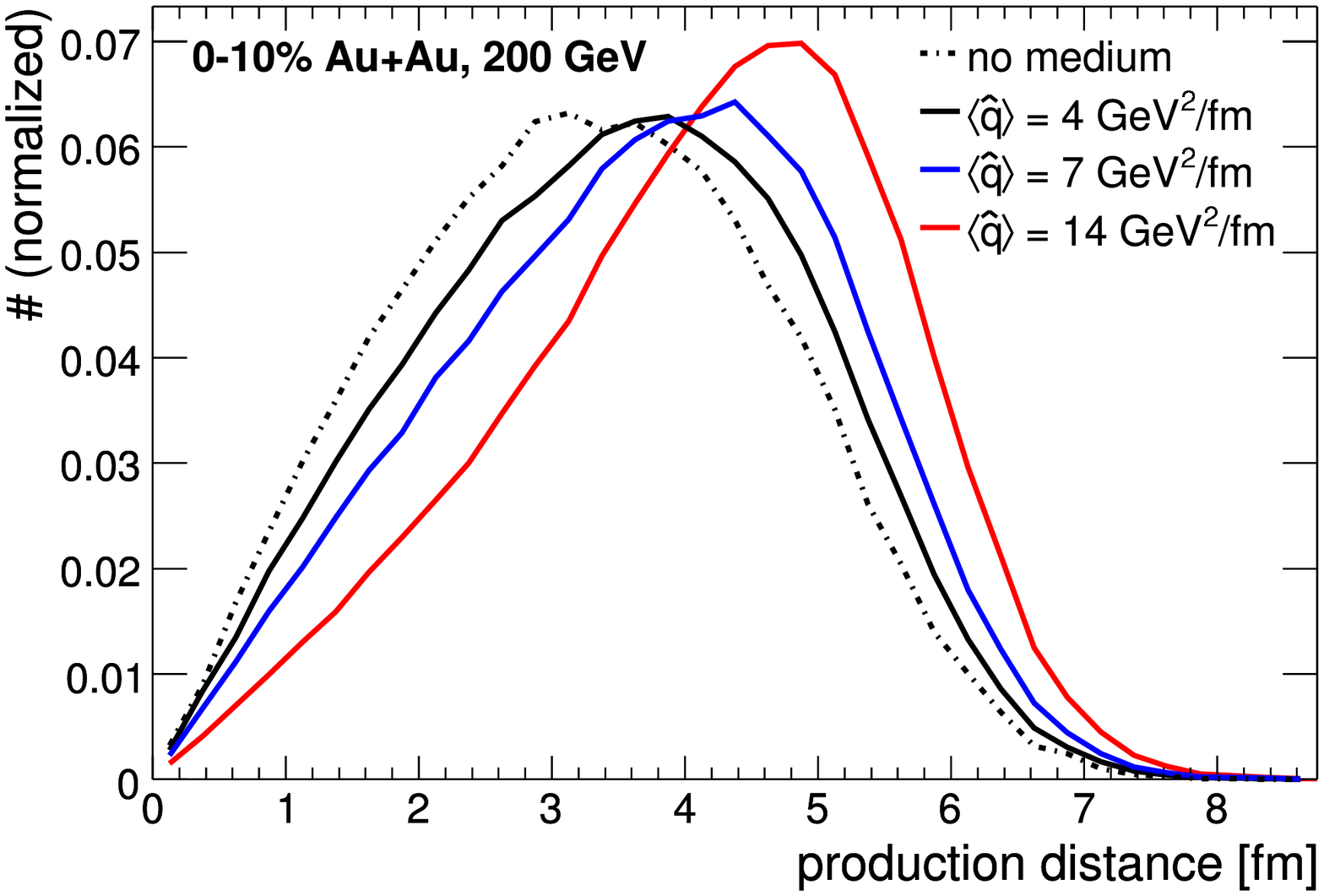}}
\subfigure[]{\label{cl:fig8c}
\includegraphics[width=0.49\textwidth]{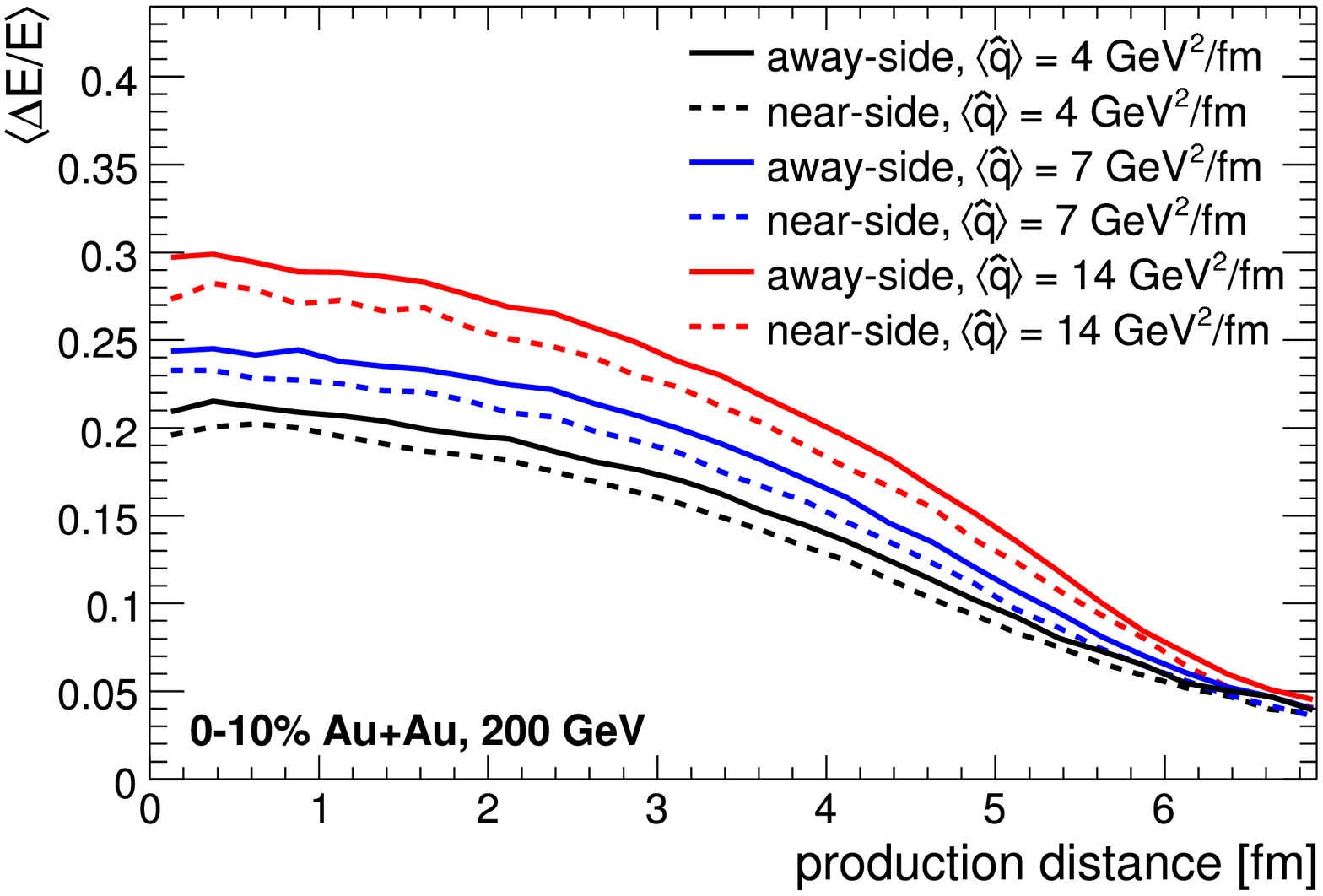}}
\end{center}
\vspace{-0.4cm}
\caption[]{\label{cl:fig8}
 \subref{cl:fig8a}~The probability of zero-energy loss for the away-side parton versus the distance 
 of the production point from the collision center, \subref{cl:fig8b} the distribution of production 
 distances from the collision center and~\subref{cl:fig8c} the average relative energy loss for near-side 
 and away-side partons versus the distance of the production point from the collision center for $\avq=4$, 
 $7$ and $14~\qun$ in 0--10\% central \AuAu\ collisions. In all cases, the trigger cuts are for
 hadron pairs within $8<\ptt\le15~\gev$ and $6~\gev\le\pta\le\ptt$.
} 
\enlargethispage{1cm}
\end{figure}

For the PQM framework, the situation is illustrated in \Fig{cl:fig7}, which
shows production points and emission direction for surviving back-to-back pairs 
in the transverse plane for 0--10\% central \AuAu\ collisions. Each parton pair 
shown yields a hadron pair within $8<\ptt\le15~\gev$ and $6~\gev\le\pta\le\ptt$.
The chosen set of transport coefficients are the same as previously used to describe 
the $\IAAa$ data, $\avq=4$, $7$ and $14~\qun$, in the non-reweighted case.
In panels of \Fig{cl:fig7}, the center of any line indicates the production point 
of a parton pair, and the two partons emerge along the line in opposite 
direction. The color of each line either indicates the medium density (left) or
the relative energy loss (right panels) of the away-side parton, which is defined 
as the parton fragmenting into the associated hadron. The left panels of \Fig{cl:fig7}
clearly illustrates that the largest medium densities are encountered by away-side 
partons that pass through the central zone of the collision, which is expected for
a static scenario. The core is surrounded by dijets that escape tangentially with 
respect to the central region~\Ref{cl:qm05}. 
In the case of an expanding medium, 
the interpretation
of the underlying trigger bias remains the same: surviving dijets are always selected 
from regions (and times) of the collision evolution where they suffered the least 
amount of interactions with the medium. The right panels of \Fig{cl:fig7}, which 
color-code the relative energy loss of the away-side parton, illustrate this point 
from a slightly different perspective. It is apparent that in all three cases,
the energy loss is close to zero for a large fraction of the dijets, even for the core 
region. This is due to the probability of zero-energy loss for the away-side parton,
$P(\Delta E^{\rm away}=0)$, which for the core is already about 0.05, even for the 
$\avq=14~\qun$ case. 
\enlargethispage{0.5cm}
As shown in \Fig{cl:fig8a}, $P(\Delta E^{\rm away}=0)$ quickly reaches values between 0.1 
and 0.3 for the $d\lsim 4~\fm$, where $d=\sqrt{x_0^2+y_0^2}$ is the distance of the
production point to the collision center. A large fraction of these parton pairs are 
initially produced, see \Fig{cl:fig8b}~(dashed line, no medium). The same figure illustrates 
the shift of the production point distribution to larger distances with increasing medium 
densities, when energy loss is included. The mean values change from $3.2~\fm$ in vacuum to 
$3.5$, $3.7$ and $4.1~\fm$ for $\avq=4$, $7$ and $14~\qun$, respectively.
In \Fig{cl:fig8c}, we finally present the average $\Delta E/E$ versus $d$ for 
near-side and away-side partons.
For distances of $d\gsim4~\fm$, that contribute significantly to the near- and
away-side yields, $0.1\lsim\Delta E/E\lsim0.25$ for the three densities, and the 
average relative energy loss is nearly the same on both sides.

\section{\label{cl:concl}Summary}
\enlargethispage{0.5cm}
Jet quenching effects at the top RHIC energy are discussed within the Parton Quenching 
Model that includes a probabilistic treatment of the BDMPS quenching weights and a 
Glauber-based implementation of the collision geometry. The available high-$\pt$ data 
for $\RAA$, $\IAAa$ and their centrality dependence constrain the extracted medium 
density to about $3\lsim\avq\lsim9~\qun$ for central \CuCu\ and $4\lsim\avq\lsim14~\qun$ 
for central \AuAu\ collisions. Our analysis suggests that particle production in central
collisions is `surface' dominated, not only for single hadrons, but also for dijets. The 
properties of surviving dijets therefore are very similar to the vacuum properties.

\begin{thebibliography}{}

\bibitem{cl:phenixRAA}
  S.S.\ Adler {\it et al.} [PHENIX], {\noit Phys.\ Rev.\ Lett.}~{\bf 91} (2003) 072301,
  {\noit Phys.\ Rev.}~{\bf C69} (2004) 034910.

\bibitem{cl:starRAA}
  J.\ Adams {\it et al.} [STAR], {\noit Phys.\ Rev.\ Lett.}~{\bf 91} (2003) 172302.

\bibitem{cl:starIAA}
  C.\ Adler {\it et al.} [STAR], {\noit Phys.\ Rev.\ Lett.}~{\bf 90} (2003) 082302,
  {\noit Phys.\ Rev.\ Lett.}~{\bf 95} (2005) 152301.

\bibitem{cl:na57RCP}
  F.\ Antinori {\it et al.} [NA57], {\noit Phys.\ Lett.\ B}~{\bf 623} (2005) 17;
  C.\ Hohne {\it et al.} [NA49], \arxiv{nucl-ex/0510049}.

\bibitem{cl:phobosRAA}
  B.\ Alver {\it et al.} [PHOBOS], {\noit Phys.\ Rev.\ Lett.}~{\bf 96}, 212301 (2006). 

\bibitem{cl:dAu}
 B.B.\ Back {\it et al.}\ [PHOBOS], {\noit Phys.\ Rev.\ Lett.}~{\bf 91} (2003) 072302;
 J.\ Adams {\it et al.} [STAR], {\noit ibidem} 072304;
 S.S.\ Adler {\it et al.} [PHENIX], {\noit ibidem} 072303;
 I.\ Arsene {\it et al.} [BRAHMS], {\noit ibidem} 072305.

\bibitem{cl:eloss}
  C.A.\ Salgado, \arxiv{hep-ph/0510062}; X.N.\ Wang, \arxiv{nucl-th/0511001};
  I.\ Vitev, \arxiv{hep-ph/0603010}.

\bibitem{cl:collloss}
  A.\ Peshier, these proceedings, \arxiv{hep-ph/0607275}.

\bibitem{cl:modelsdata}
  D.\ d'Enterria, {\noit Eur.\ Phys.\ J.\ C}~{\bf 43}, 295 (2005). 

\bibitem{cl:PQM}
  A.\ Dainese, C.\ Loizides and G.\ Pai\'c, {\noit Eur.\ Phys.\ J.}~{\bf C38} (2005) 461. 

\bibitem{cl:carlosurs}
  C.A.\ Salgado and U.A.\ Wiedemann, {\noit Phys.\ Rev.}~{\bf D68} (2003) 014008.

\bibitem{cl:photons}
  S.\ Adler {\it et al.} [PHENIX], {\noit Phys.\ Rev.\ Lett.}~{\bf 94}, 232301 (2005).

\bibitem{cl:phenixPrelimRAA}
  M.\ Shimomura [PHENIX], \arxiv{nucl-ex/0510023}.

\bibitem{cl:preQMprediction}
  C.\ Loizides, QM'05 poster (not published), see
  P.~M.~Jacobs and M.~van Leeuwen, \arxiv{nucl-ex/0511013}.

\bibitem{cl:baier}
  R.\ Baier, {\noit Nucl.\ Phys.}~{\bf A715} (2003) 209.

\bibitem{cl:ekrt}
  K.J.~Eskola {\it et al.}, {\noit Nucl.\ Phys.}~{\bf B570} (2000) 379.

\bibitem{cl:wang}
  E.\ Wang and X.N.\ Wang, {\noit Phys.\ Rev.\ Lett.}~{\bf 89} (2002) 162301.

\bibitem{cl:drees}
  A.\ Drees, H.\ Feng and J.\ Jia, {\noit Phys.\ Rev.\ C}~{\bf 71}, 034909 (2005).

\bibitem{cl:qm05}
  A.\ Dainese, C.\ Loizides and G.\ Paic, \arxiv{hep-ph/0511045}. 

\bibitem{cl:starDirectJets}
  J.~Adams {\it et al.} [STAR], \arxiv{nucl-ex/0604018}.

\bibitem{cl:fragile}
  K.J.\ Eskola {\noit et al.}, {\noit Nucl.\ Phys.}~{\bf A747} (2005) 511.

\end{thebibliography}
\end{document}